\documentclass[%
 reprint,
 showpacs,preprintnumbers,
 amsmath,amssymb,
 aps,
 prb,
]{revtex4-1}
\usepackage{dcolumn}
\usepackage[dvipdfmx]{graphicx}
\usepackage{subfigure}
\usepackage[dvipdfmx]{color}
\usepackage{mathrsfs}

\makeatletter
\def\btt#1{\texttt{\@backslashchar#1}}
\DeclareRobustCommand\bblash{\btt{\@backslashchar}} \makeatother

\begin{document}

\title{Spin and electric currents induced by a spin-motive force in two-dimensional Dirac semimetals protected by nonsymmorphic symmetries}

\author{Tetsuro Habe}
\affiliation{Department of Physics, Osaka University, Toyonaka, Osaka 560-0043, Japan}

\date{\today}

\begin{abstract}
We theoretically study the spin and charge currents induced by a spin-motive force in a two-dimensional Dirac semimetal protected by nonsymmorphic symmetries.
Glide mirror plane symmetry, a nonsymmorphic symmetry, leads to a constraint to the induced current; the spin-motive force acting on out-of-plane spin (in-plane spin) induces the pure spin (charge) current.
We calculate the response function to the spin-motive force in linear response theory and find that the conductivity for the pure spin current remains non-zero even if the Fermi energy is crossing the node of linear dispersion.
We also find that the dissipationless spin current is induced at the charge neutral point. 
\end{abstract}

\pacs{73.22.-f}

\maketitle
\section{Introduction}
Massless Dirac fermionic states of electrons in two-dimensional (2D) systems have shown several kinds of fascinating transport phenomena; Klein tunneling\cite{Katsnelson2006}, the Hall effect induced by a external field coupling to some discrete degrees of freedom\cite{Gusynin2005,Qi2006,Xiao2007,Liu2008,Zu-Chang2013}, the spin and valley Hall effects\cite{Bernevig2006,Konig2007,Xiao2007}, and specular Andreev Reflection.\cite{Beenakker2006,Efetov2016}
These phenomena are attributed to the variation of spin (pseudo-spin) degree of freedom with respect to the wave vector and the linear energy dispersion of  the electronic states.\cite{Murakami2007,Hsieh2009}
There have been two famous examples of such a two-dimensional electronic system, graphene\cite{Neto2009} and the surface electronic states of three-dimensional topological insulators.\cite{Fu2007}

Recently, the nonsymmorphic symmetry-protected (NSSP) Dirac semimetal was proposed by Young and Kane\cite{Young2015}, and the electronic states in the semimetal have linear energy dispersion and they are doubly degenerated at each wave vector as long as time-reversal and nonsymmorphic symmetries are preserved.
We find that a coexistence of glide mirror plane symmetry, a nonsymmorphic symmetry, and the linear energy dispersion leads to unconventional responses to a spin-motive force, which is defined in Ref.\ \onlinecite{Shi2006}.

In this paper, we investigate the spin and charge currents induced by a spin-motive force, e.g. a slanting Zeeman field in experiments\cite{Tokura2006,Pioro2008} , in the NSSP 2D Dirac semimetal.
In Sec.\ \ref{Sec_electronic_states}, we consider the proper model of the NSSP 2D Dirac semimetal and give the charge and spin current operators. 
In Sec.\ \ref{Sec_response_function}, we discuss the role of glide mirror plane symmetry in the current induced by the spin-motive force, and find that the spin motive force acting on the out-of-plane spin and in-plane spin induce the pure spin current and the charge current, respectively, because of the symmetry.
We also calculate the response function to the fields in linear response theory and find the spin-spin conductivity for the out-of-plane field remains non-zero even if the Fermi energy crosses the node of Dirac cone.

\section{Electronic states in NSSP 2D Dirac semimetal}\label{Sec_electronic_states}

We consider the electronic states in the NSSP 2D Dirac semimetal and they can be described by \cite{Young2015},
\begin{align}
H_0=&2t\tau_x\cos\frac{k_x}{2}\cos\frac{k_y}{2}+t_2(\cos k_x+\cos k_y)\nonumber\\
&+t_{so}\tau_z(\sigma_y\sin k_x-\sigma_x\sin k_y)+\Delta_1\sin\frac{k_x}{2}\sin\frac{k_y}{2}\tau_x,\label{Hamiltonian}
\end{align}
where $\tau$ and $\sigma$ are Pauli matrices in the sublattice and spin spaces.
Here, $t$ and $t_2$ are the nearest and next-nearest hopping matrices, respectively.
The spin-orbit interaction is represented by $t_{so}$, and $\Delta_1$ is induced by a deformation of the lattice structure.
One candidate simulated by this Hamiltonian is iridium oxide superlattce\cite{Chen2014}.

\subsection{Low-energy states}

The electronic band structure has two Dirac points appearing at the symmetrical points $X_1=(\pi,0)$ and $X_2=(0,\pi)$, and the dispersion is a linear function of the relative wave vector with respect to the points.
Around the Dirac points, the electronic states can be described  by the $2\times2$ effective Hamiltonian for each eigenvalue of the glide mirror plane operator $\xi_z=\tau_x\sigma_z$ with $s_z=\tau_x$\cite{Habe2017},
\begin{align}
H_{\zeta}=(u_{\zeta}p_x+u'_{\zeta}p_y)s_z-\zeta v_{so}(\xi_zs_yp_x+s_xp_y),\label{Effective_Hamiltonian}
\end{align}
where $\boldsymbol{p}$ is the relative wave number with respect to the Dirac point, and the valley index $\zeta=1$ ($\zeta=-1$) represents the valley around the Dirac point of $X_1$ ($X_2$). 
The velocities are defined by $v_{so}=t_{so}a/\hbar$, and $(u_{\zeta},u'_{\zeta})=(a/\hbar)(t,-\Delta_1)$ for $\zeta=1$ and $(a/\hbar)(-\Delta_1,t)$ for $\zeta=-1$ with the lattice constant $a$.
Here, we ignore the second nearest neighbor hopping and the correction by it is discussed in the following section. The energy dispersion is particle-hole symmetric $\varepsilon=\pm\varepsilon_0$ and depending on the direction of wave vector $\boldsymbol{p}=(p\cos\theta,p\sin\theta)$,
\begin{align}
\varepsilon_0=v_\theta p,
\end{align}
where the Fermi velocity $v_\theta$ is given by $v_\theta=\sqrt{\bar{u}^2\cos^2(\theta-\theta_\zeta)+v_{so}^2}$ with $(\bar{u}\cos\theta_\zeta,\bar{u}\sin\theta_\zeta)=(u_\zeta,u_\zeta')$.

\begin{figure}[htbp]
\begin{center}
 \includegraphics[width=70mm]{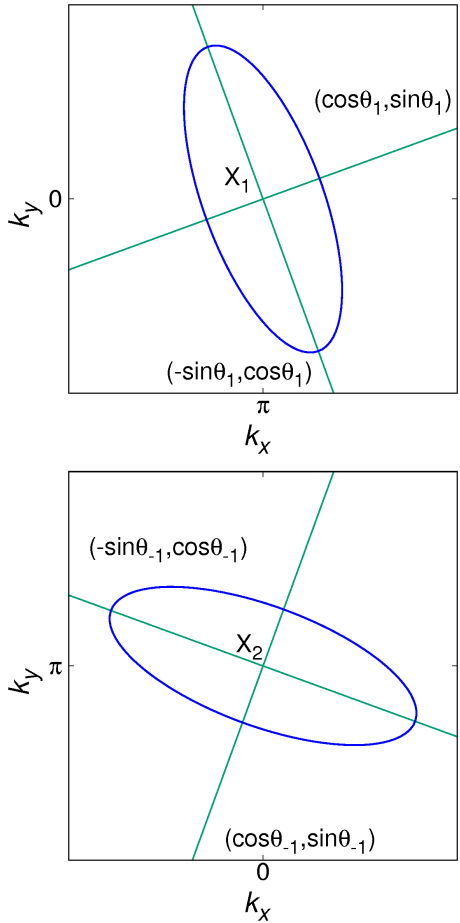}
\caption{The schematic picture of the Fermi surface around the $X_1$ and $X_2$.
 }\label{Fig_Fermi_surface}
\end{center}
\end{figure}
The Fermi surface is strongly warped because of the anisotropic Fermi velocity, and it is represented by $p(\theta)=\varepsilon_F/v_\theta$ as shown in Fig.\ \ref{Fig_Fermi_surface}.
The symmetrical axis of the Fermi surface is tilted by the lattice distortion for $u_\zeta u_\zeta'\neq0$ from the $p_x$ and $p_y$ axises where the short axis and long axis are along $(\cos\theta_\zeta,\sin\theta_\zeta)$ and $(-\sin\theta_\zeta,\cos\theta_\zeta)$, respectively.
In the Fermi surface, the direction of the pseudo-spin $\boldsymbol{s}$ i.e., the vector component $|\theta,\xi_z,s\rangle$ of the eigenstate, changes with the direction of the wave vector $\theta$, the glide mirror plane parity $\xi_z$, and the band index $s$.
Here, $s$ indicates the upper cone for $s=+$ and the lower cone for $s=-$.

\subsection{Charge and spin current}

We consider the flows of charge and spin of electrons as charge and spin currents, and discuss the representation of them as operators in the 2D Dirac semimetal.
The charge current can be defined by a flow of electrons in the massless fermionic states with the velocity given by $v_\nu=\partial H_\zeta/\partial p_\nu$, and the charge current operator can be represented by $j^e_\mu=ev_\mu$ with the electronic charge $e$.
The velocity operator can be obtained from Eq.\ (\ref{Effective_Hamiltonian}) as 
\begin{align}
v_x=u_\zeta s_z-\zeta\xi_z v_{so}s_y,\;\;v_y=u_\zeta's_z-\zeta v_{so}s_x,\label{def_charge_current}
\end{align} 
and thus the charge current operator is invariant under the glide mirror plane operation.

The definition of spin current is more complicated in general case where the spin is not conserved quantity and the time-derivative of spin is equal to the conventional spin current $j_{\mu}^s=\{\sigma_z,v_\mu\}/2$ plus the spin-torque $\boldsymbol{r}d\sigma_z/dt$\cite{Shi2006}.
The conventional spin current operator describes the difference of electronic flows with opposite spins, and the operator for the out-of-plane spin current is given by
\begin{align}
j^s_{x}=\frac{u_\zeta}{2}\xi_z,\;\;j^s_{y}=\frac{u_\zeta'}{2}\xi_z.\label{def_spin_current}
\end{align}
Here, $j^s_{\mu}$ is a conserved current in the Dirac semimetal because of $[j^s_{\mu},H]=0$, and thus we can separately discuss the contributions of the conventional spin current and spin torque to the spin current.%
The local spin torque appears in the absence of inversion symmetry and presence of spin-orbit interaction\cite{Berger1996,Berger1997,Zhou2008}, however the net spin transfer by the torque vanishes on average in the bulk\cite{Shi2006}. 
Therefore, we can discuss the spin current by the conventional spin current in Eq.\ (\ref{def_spin_current}).

\section{Response function to a spin-motive force}\label{Sec_response_function}

We consider the charge and spin currents induced by a spin-motive force in linear response theory, and discuss the constraints of glide mirror plane symmetry to spin and flow directions of the induced current.
The effect of such a force can be represented by a perturbation of the coupling between a spin-motive force $F_{\mu\nu}=\partial B_\mu/\partial r_\nu$, which can be obtained from the Zeeman field $\boldsymbol{B}$, and the spin-displacement $d_{\mu\nu}^s=\sigma_\mu r_\nu$ with the displacement $r_\nu$ in the $\nu$ direction.\cite{Zutic2002,Fabian2002,Shi2006}
In linear response theory, the conductivity for charge and spin currents can be defined by the mean current over the spin-motive force,
\begin{align}
\sigma^{qs}_{\mu\nu\rho}=\langle j^q_\mu\rangle/F_{\rho\nu},
\end{align}
where $\langle j^q\rangle$ is the expectation value of the current of charge $j^c$ or spin $j^s$.
At zero temperature, the dc conductivity is given by
\begin{align}
\sigma^{qs}_{\mu\nu\rho}=&\int\frac{dpd\theta}{(2\pi)^2}p\sum_{\xi_z,\xi_z'}\mathrm{Re}[\langle \theta,\xi_z,+|j_\mu^q|\theta,\xi_z',+\rangle\notag\\&\times\langle\theta,\xi_z',+|d^s_{\rho\nu}|\theta,\xi_z,+\rangle]\delta(v_\theta p-\varepsilon_F),\label{eq_spin_conductivity}
\end{align}
with a Fermi energy $0<\varepsilon_F$, where the electronic eigenstates of Eq.\ (\ref{Effective_Hamiltonian}) are represented by $|\theta,\xi_z,+\rangle$ in the upper cone.\cite{Habe2017}
Here, the spin displacement operator is represented by the long-range part through the unit cells and the short-range part between the sublattices,
\begin{align}
\begin{aligned}
d_{x\nu}^s=&\xi_x\frac{1}{i}\frac{\partial}{\partial p_\nu}+d_\nu s_x\\d_{y\nu}^s=&s_z\xi_y\frac{1}{i}\frac{\partial}{\partial p_\nu}+d_\nu s_y\xi_z\\d_{z\nu}^s=&s_z\xi_z\frac{1}{i}\frac{\partial}{\partial p_\nu}-d_\nu s_y\xi_y,
\end{aligned}\label{def_spin_displacement}
\end{align}
where $d_\nu$ is the $\nu$ component of the relative vector between the two sublattices.

\subsection{Constraints by glide mirror plane symmetry}\label{discuss_symmetry}

The conductivity tensor in Eq.\ (\ref{eq_spin_conductivity}), in general, represents several types of conductivity corresponding to the indexes, but glide mirror plane symmetry leads to constraints to non-zero conductivity.
The electronic states with different glide mirror plane parity $\xi_z=\pm1$ at each wave vector can be transformed by $|\theta,-\xi_z,s\rangle=\xi_\mu|\theta,\xi_z,s\rangle$ for $\mu=x$ or $y$ plus $s_y\rightarrow-s_y$, and thus the spin displacement and current operators must be invariant under this transformation, i.e., operators in Eq.\ (\ref{eq_spin_conductivity}) are independent of $\xi_x$ and $\xi_y$.
Moreover, the contributions from degenerated cones with different $\xi_z$ cancel each other out when the net sign change of $j^q$ and $d^s$ under $\xi_z\rightarrow-\xi_z$ and $s_y\rightarrow-s_y$.
As a consequence of the symmetry, we can conclude that the charge (spin) current is never induced by the spin-motive force for out-of-plane (in-plane) spin, 
\begin{align}
\begin{aligned}
\sigma^{es}_{\mu\nu z}&=0,\\
\sigma^{ss}_{\mu\nu x}&=\sigma^{ss}_{\mu\nu y}=0,
\end{aligned}\label{eq_symmetrical_constraints}
\end{align}
even if we consider the higher order term of the relative wave number $p$ around the Dirac points.

\subsection{Spin-motive force acting on out-of-plane spin}\label{Out-of-plane_spin}

\begin{figure}[htbp]
\begin{center}
 \includegraphics[width=70mm]{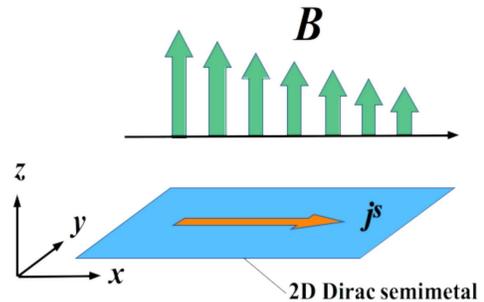}
\caption{Schematic picture of the spin-motive force in the out-of-plane direction.
 }\label{Fig_out-of-plane}
\end{center}
\end{figure}
We calculate the spin current induced by a spin-motive force acting on an out-of-plane spin $\sigma_z$.
At the valley $\zeta$, the spin-spin conductivity is given by
\begin{align}
\sigma_{\mu\nu z}^{ss}(\zeta)=\frac{\hbar}{2\pi}\left(1-\frac{v_{so}/\bar{u}}{\sqrt{v_{so}^2/\bar{u}^2+1}}\right)c^{ss}_{\mu\nu}(\theta_\zeta),\label{eq_zspin_zspin_conductivity}
\end{align}
where the relation between the flow direction and the field direction is described by a $2\times2$ matrix $c^{s_z}_{\mu\nu}$,
\begin{align}
\begin{pmatrix}
c^{ss}_{xx}(\theta_\zeta)&c^{ss}_{xy}(\theta_\zeta)\\
c^{ss}_{yx}(\theta_\zeta)&c^{ss}_{yy}(\theta_\zeta)
\end{pmatrix}=\begin{pmatrix}
\cos^2\theta_\zeta&\cos\theta_\zeta\sin\theta_\zeta\\
\cos\theta_\zeta\sin\theta_\zeta&\sin^2\theta_\zeta
\end{pmatrix}.\label{eq_angle_spin_spin_conductivity}
\end{align}
The flow direction is parallel to the short axis $(\cos\theta_\zeta,\sin\theta_\zeta)$ of the Fermi surface at the $X_{\zeta}$ in Fig.\ \ref{Fig_Fermi_surface}, and the current is proportional to the projection of the spin-motive force to the same axis.

The directivity of spin flow in Eq.\ (\ref{eq_angle_spin_spin_conductivity}) is attributed to the anisotropy of the nearest-neighbor hopping term in two valleys.
Near the Dirac points, both the nearest-neighbor hopping and the spin-orbit interaction contribute to the electron transfer, but the spin transfer is governed by the nearest-neighbor hopping.
Thus, the electronic spin flows in the direction parallel to the short axis of the Fermi surface in each valley, and the electronic spin flows in different directions around the two Dirac points.
Therefore, the flow direction of the net spin current is not restricted in the 2D Dirac semimetal with a realistic lattice deformation of $\theta_\zeta\ll\pi/4$.

The spin-spin conductivity is nearly unity under the condition of $v_{so}/\bar{u}\ll1$, i.e. the nearest neighbor hopping matrix is much larger than the spin-orbit coupling, and independent of the Fermi energy.
The independence of the Fermi energy is attributed to the dimension of the spin-motive force, the first term of $d^s_{z\nu}$ in Eq.\ (\ref{def_spin_displacement}), and the linear dispersion of the electronic states.
The dimension of the spin-motive force is $T^{-1}L^{-1}$, where $T$ and $L$ represent the dimension of time and length, and that of current density is also $T^{-1}L^{-1}$ except for the dimension of charge or spin.
Thus, the spin-spin conductivity has the dimension of charge or spin and it is independent of the Fermi energy because there is no particular energy scale in the electronic system with a linear dispersion where the short range part of spin displacement in Eq.\ (\ref{eq_symmetrical_constraints}), proportional to $d_\nu$, does not associated with the current due to glide mirror plane symmetry as discussed in Sec.\ \ref{discuss_symmetry}.

\subsection{Spin-motive force acting on in-plane spin}
\begin{figure}[htbp]
\begin{center}
 \includegraphics[width=70mm]{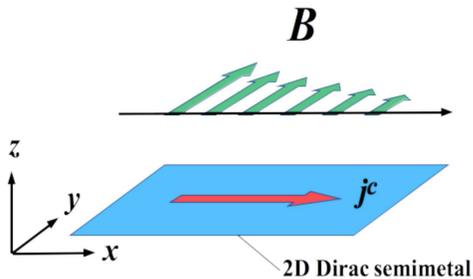}
\caption{Schematic picture of the spin-motive force in the in-plane direction.
 }\label{Fig_in-plane}
\end{center}
\end{figure}
The spin-motive force for in-plane spin leads to charge current, and the spin flow is absent according to the discussion in Sec.\ \ref{discuss_symmetry}.
In the $\zeta$ valley, the charge-spin conductivity, the response function of the charge current to the spin-motive force, is given by
\begin{align}
\sigma^{es}_{\mu\nu x}(\zeta)=2\frac{e}{h}\frac{\zeta\varepsilon_F}{\sqrt{v_{so}^2+\bar{u}^2}}g^{s_x}_{\mu}d_\nu\label{eq_charge_xspin_conductivity}\\
\sigma^{es}_{\mu\nu y}(\zeta)=2\frac{e}{h}\frac{\zeta\varepsilon_F}{\sqrt{v_{so}^2+\bar{u}^2}}g^{s_y}_{\mu}d_\nu\label{eq_charge_yspin_conductivity}
\end{align}
where the vector indicating the current direction is represented by
\begin{align}
\boldsymbol{g}^{s_x}=&
\begin{pmatrix}
\cos\theta_\zeta\sin\theta_\zeta/(v_{so}^2/\bar{u}^2+1))\\
-\cos^2\theta_\zeta-\sin^2\theta_\zeta/(1+\bar{u}^2/v_{so}^2)
\end{pmatrix}\\
\boldsymbol{g}^{s_y}=&
\begin{pmatrix}
-\sin^2\theta_\zeta-\cos^2\theta_\zeta/(1+\bar{u}^2/v_{so}^2)\\
\cos\theta_\zeta\sin\theta_\zeta/(v_{so}^2/\bar{u}^2+1)\\
\end{pmatrix}.
\end{align}
When the spin-orbit interaction is much smaller than the hopping matrix $v_{so}/\bar{u}\ll1$, the charge flow direction is nearly parallel to the long axis $(-\sin\theta_\zeta,\cos\theta_\zeta)$ of the Fermi surface in Fig.\ \ref{Fig_Fermi_surface}.

The charge-spin conductivity is quite different from the spin-spin conductivity in the dependence on the Fermi energy in Eq.\ (\ref{eq_charge_xspin_conductivity}) and (\ref{eq_charge_yspin_conductivity}).
This is because the charge current is correlated with the spin-displacement between the sublattices in Eq.\ (\ref{def_spin_displacement}).
The displacement has a typical length scale $d_\nu$ but the electronic states with the linear dispersion have no particular length without the inverse of the Fermi wave number, and thus the conductivity must depends on the Fermi energy.
The Fermi energy dependence means that one can obtain the larger charge-spin conductivity by increasing the charge density.

\subsection{Contribution of second-nearest neighbor hopping}
In this subsection, we discuss the contribution from the second nearest neighbor hopping with $t_2$ in Eq.\ (\ref{Hamiltonian}) to the conductivities.
The second nearest-neighbor hopping corrects the conductivity obtained in previous subsections but it does not change zero components of Eq.\ (\ref{eq_spin_conductivity}).
This is because the symmetrical property of charge and spin current operators, discussed in Sec.\ \ref{discuss_symmetry}, is unchanged from that of Eq.\ (\ref{def_charge_current}) and (\ref{def_spin_current}).

We evaluate the correction to the conductivities, which is driven from the second nearest-neighbor hopping, by calculating it in linear response theory.
The exact form of the correction to the spin-spin conductivity in Eq.\ (\ref{eq_zspin_zspin_conductivity}) includes the elliptic integrals but the upper limit is given by $\Delta\sigma^{ss}<2\pi\hbar t_2\varepsilon_F/( t_{so}\sqrt{t^2+{\Delta_1}^2})$.
The correction to the charge-spin conductivity in Eq.\ (\ref{eq_charge_yspin_conductivity}) is also given by $\Delta \sigma^{es}\simeq (t_2\varepsilon_F/{t_{so}}^2)\sigma^{es}$.
In both caseses, the correction by the second nearest-neighbor hopping reduces with the decrease in the charge density.
Thus, the correction by the nearest neighbor hopping is qualitatively small in the low carrier density, but the spin current associated with this hopping is a small dissipative component.
Therefore, the completely dissipationless spin current is realized at the charge neutral point.

\section{Conclusion}

In conclusion, we study the spin and charge currents induced by a spin-motive force in a 2D Dirac semimetal protected by glide mirror plane symmetry, a nonsymmorphic symmetry.
Glide mirror plane symmetry provides the conservation of spin current and the pure spin (charge) current in the presence of the spin-motive force acting on out-of-plane (in-plane) spin.
We calculate the response function to the force in linear response theory and give the analytic formulations.
We find that the spin-spin conductivity remains non-zero and the dissipationless spin current is obtained at the charge neutral point.

\bibliography{2D_Dirac}

\end{document}